O. Sergijenko
Astronomical Observatory of
Taras Shevchenko National University of Kyiv

# Dipole of the luminosity distance as a test for dark energy models

*The dependence of Hubble parameter on redshift can be determined directly from the dipole of luminosity distance to Supernovae Ia. We investigate the possibility of using the data on dipole of the luminosity distance obtained from the Supernovae Ia compilations SDSS, Union2.1, JLA and Pantheon to distinguish the dark energy models.*

*Introduction.*
According to the Planck data the dark energy in current epoch is close to the cosmological constant. This makes it much harder to detect the temporal variation of the dark energy equation of state parameter and to distinguish the dark energy models. The reliability of *w(z)* determination will be significantly higher if the dependence *H(z)* is measured directly instead of (or along with) the luminosity or angular diameter distances since taking numerical derivatives from the current observational data leads to the inaccurate results. Recent interest to testing the isotropy of the Supernovae Ia magnitude-redshift relation [1,2] brings attention to the dipole of luminosity distance as a possible direct measure of the Hubble parameter [3]. In this paper we make an attempt to re-estimate the potential of the luminosity distance dipole to discriminate the dark energy models using current data on Supernovae Ia.

*Dipole of the luminosity distance.*
According to [3,4] the first-order expansion of directional dependence of the luminosity distance reads:

$$d_L(z,\mathbf{n}) = d_L^{(0)}(z) + d_L^{(1)}(z)(\mathbf{n}\cdot\mathbf{e}),$$

where the monopole is

$$d_L^{(0)}(z) = (1+z)\int_0^z \frac{dz'}{H(z')}$$

and the dipole is

$$d_L^{(1)}(z) = \frac{|\mathbf{v}_0|(1+z)^2}{H(z)}.$$

The variance of dipole

$$\left(\Delta d_L^{(1)}(z)\right)^2 = 3\left(\frac{\ln(10)}{5}\right)^2 \Delta m^2 \left(d_L^{(0)}(z)\right)^2$$

leads to the following estimate for precision of the *H(z)* determination [3]:

$$\frac{\Delta H(z)}{H(z)} = \frac{\sqrt{3}\ln(10)}{5|\mathbf{v}_0|}\frac{d_L^{(0)}(z)H(z)}{(1+z)^2}\Delta m.$$

*Model and data.*
As the dark energy model we use the minimally coupled classical scalar field with the equation of state parameter obtained from the condition $c_a^2 = \frac{\dot{p}_{de}}{\dot{\rho}_{de}} = const$ [5]:

$$w(a) = \frac{p_{de}}{\rho_{de}} = \frac{(1+c_a^2)(1+w_0)}{1+w_0-(w_0-c_a^2)a^{3(1+c_a^2)}} - 1.$$

We investigate 2 cases: distinguishing between the best-fit quintessential and phantom models with the parameters obtained from the same dataset (as it was done in [5,6]) and distinguishing the mean model from the model with all parameters at 1σ or 2σ confidence limits (in the manner of [7]).

The cosmological parameters and their confidence ranges are obtained by the Monte Carlo Markov chain (MCMC) [8] method implemented in the CosmoMC code (http://cosmologist.info/cosmomc). We assume the spatial flatness of the Universe.

In the first case we use the bets-fit parameters (Table 1) estimated in [5] from the following datasets:
- CMB temperature fluctuations and polarization angular power spectra from the 7-year WMAP data (WMAP7)[9-11];
- Baryon acoustic oscillations from SDSS DR7 (BAO) [12];
- Hubble constant measurements from HST (HST) [13];
- Big Bang Nucleosynthesis prior on the baryon abundance (BBN) [14,15];
- Supernovae Ia from SDSS compilation (SN SDSS) [16] (SALT2 [17] and MLCS2k2 [18] light curve fittings).

In the second case we determine the mean values of cosmological parameters and their confidence ranges (Table 2) from the combined dataset including:
- CMB TT, TE, EE angular power spectra and lensing from Planck [19];
- B-mode polarization for 2 frequency channels from BICEP2/Keck Array (BK) [20];
- power spectrum of galaxies from WiggleZ Dark Energy Survey [21];
- Supernovae Ia from JLA compilation [22];

- Hubble constant determination [23].

Here we apply flat priors with ranges of values [-2,-0.33] for $w_0$ and [-2,0] for $c_a^2$, so the dark energy model involves both quintessence and phantom subclasses.

Table 1. The best-fit values of cosmological parameters for the models with quintessential (QSF) and phantom (PSF) scalar fields determined from 2 observational datasets: WMAP7+HST+BBN+BAO+SN SDSS SALT2 ($q_1$, $p_1$) and WMAP7+HST+BBN+BAO+SN SDSS MLCS2k2 ($q_2$, $p_2$) (from [5]).

| Parameters | QSF, SALT2 ($q_1$) | PSF, SALT2 ($p_1$) | QSF, MLCS2k2 ($q_2$) | PSF, MLCS2k2 ($p_2$) |
|---|---|---|---|---|
| $\Omega_{de}$ | 0.730 | 0.723 | 0.702 | 0.692 |
| $w_0$ | -0.996 | -1.043 | -0.830 | -1.002 |
| $c_a^2$ | -0.022 | -1.120 | -0.880 | -1.190 |
| $10\Omega_b h^2$ | 0.226 | 0.223 | 0.226 | 0.223 |
| $\Omega_{cdm} h^2$ | 0.110 | 0.115 | 0.108 | 0.119 |
| $H_0$ | 70.2 | 70.4 | 66.3 | 67.8 |

Table 2. The mean values, 1σ and 2σ confidence limits for cosmological parameters obtained from the observational dataset Planck2015+WiggleZ+SN JLA+BK.

| Parameters | mean$\pm 1\sigma \pm 2\sigma$ |
|---|---|
| $\Omega_{de}$ | $0.691^{+0.012}_{-0.012}\ ^{+0.022}_{-0.024}$ |
| $w_0$ | $-1.024^{+0.062}_{-0.058}\ ^{+0.120}_{-0.125}$ |
| $c_a^2$ | $-1.460^{+0.145}_{-0.540}\ ^{+0.781}_{-0.540}$ |
| $10\Omega_b h^2$ | $0.222^{+0.002}_{-0.002}\ ^{+0.003}_{-0.003}$ |
| $\Omega_{cdm} h^2$ | $0.119^{+0.001}_{-0.001}\ ^{+0.003}_{-0.003}$ |
| $H_0$ | $67.9^{+1.2}_{-1.2}\ ^{+2.4}_{-2.3}$ |

For estimates based on the luminosity distance dipole we use the following Supernovae Ia compilations:
- SDSS [16]: 288 SNe (MLCS2k2, SALT2 light curve fitters): only statistical uncertainties;
- Union2.1 [24]: 580 SNe (SALT2): both statistical and systematic uncertainties;
- JLA [22]: 740 SNe (SALT2): both statistical and systematic uncertainties;
- Pantheon [25]: 1048 SNe (SALT2): both statistical and systematic uncertainties.

We assume $v_0$=369.0 km/s (from the CMB dipole which is due to the same motion)[26].

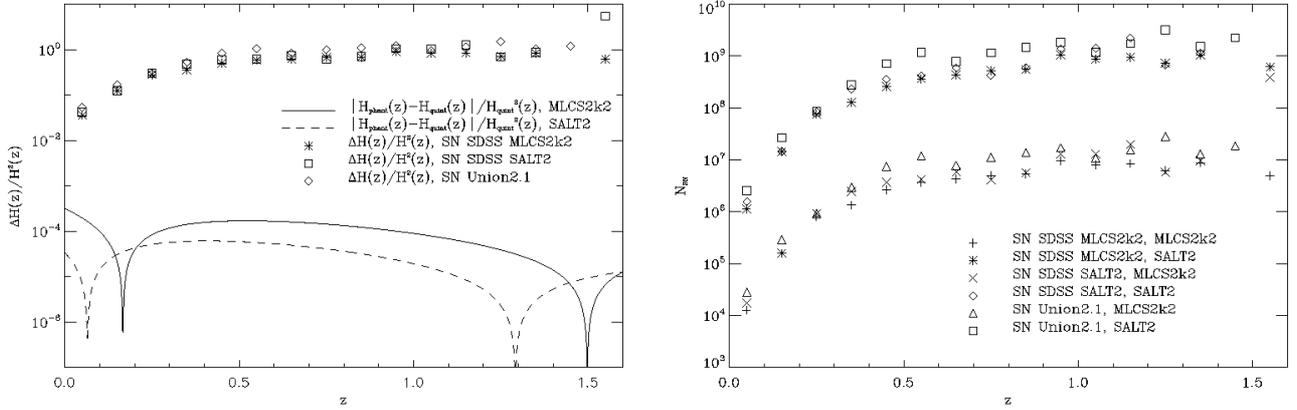

Fig. 1. Left: the theoretical relative differences $\Delta H_{model}(z)/H^2(z) \equiv |H_{phant}(z)-H_{quint}(z)|/H_{quint}^2(z)$ compared to $\Delta H(z)/H^2(z)$ from Supernovae compilations. Right: the minimal number of Supernovae that is necessary for distinguishing the models in left panel if the uncertainties of Supernovae magnitudes are the same as in the compilation from legend. After a comma we quote the data (type of Supernovae light curve fitting) used to estimate the best-fit parameters for the pair of compared models.

*Results and discussion.*
In the left panels of Fig. 1-4 we present the calculated quantities $\Delta H_{model}(z)/H^2(z)$ and compare them with the corresponding quantities $\Delta H(z)/H^2(z)$ obtained from the luminosity distance dipole using the data from Supernovae compilations in 16 redshift bins with the width 0.1 (0<*z*<1.6). In the right panels we estimate the number of Supernovae that is needed to distinguish between the models from the left panels.

From Fig. 1-4 it is clear that distinguishing between the best-fit models and between the mean model and the model with all parameters at 1σ limits is not realistic at all. The number of Supernovae necessary to distinguish the model with mean parameters from the model with all parameters at the limits of their 2σ confidence ranges is minimal in the first redshift bin (0<*z*<0.1). For SN SDSS with MLCS2k2 fitting it is 1998 or 4063, for SN SDSS with SALT2 fitting 2735 or 5563, for SN Union2.1 4466 or 9083, for SN JLA 5411 or 11006, for SN Pantheon C11 3074 or 6252 and for SN Pantheon G10 3040 or 6183 for the upper or lower limits correspondingly. For higher redshift bins the needed numbers of Supernovae are larger at least by one order of magnitude.

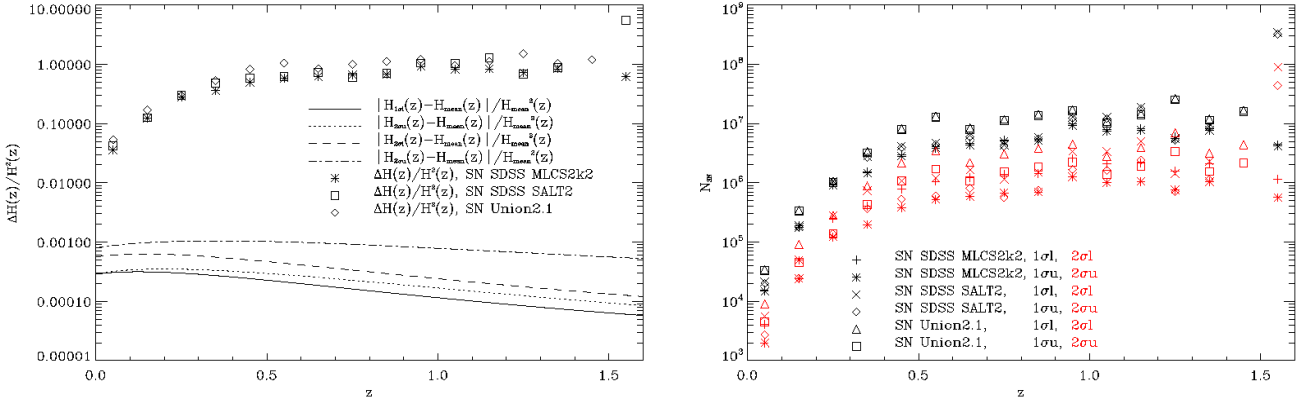

Fig. 2. Left: the theoretical relative differences $\Delta H_{model}(z)/H^2(z) \equiv |H_{1(2)\sigma}(z)-H_{mean}(z)|/H_{mean}^2(z)$ (for upper and lower limits) compared to $\Delta H(z)/H^2(z)$ from Supernovae compilations. Right: the minimal number of Supernovae that is necessary for distinguishing the models in left panel if the uncertainties of Supernovae magnitudes are the same as in the compilation from legend.

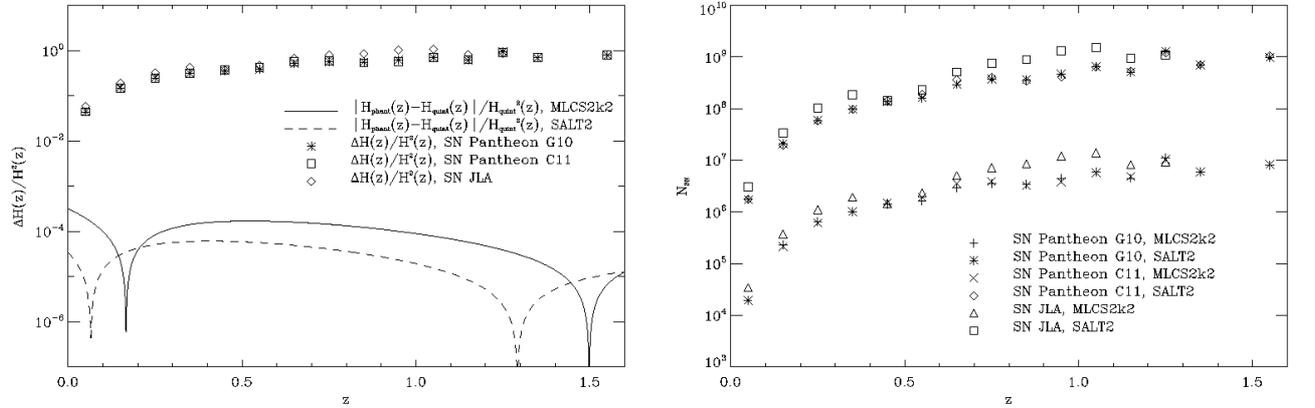

Fig. 3. Left: the theoretical relative differences $\Delta H_{model}(z)/H^2(z) \equiv |H_{phant}(z)-H_{quint}(z)|/H_{quint}^2(z)$ compared to $\Delta H(z)/H^2(z)$ from Supernovae compilations. Right: the minimal number of Supernovae that is necessary for distinguishing the models in left panel if the uncertainties of Supernovae magnitudes are the same as in the compilation from legend. After a comma we quote the data (type of Supernovae light curve fitting) used to estimate the best-fit parameters for the pair of compared models.

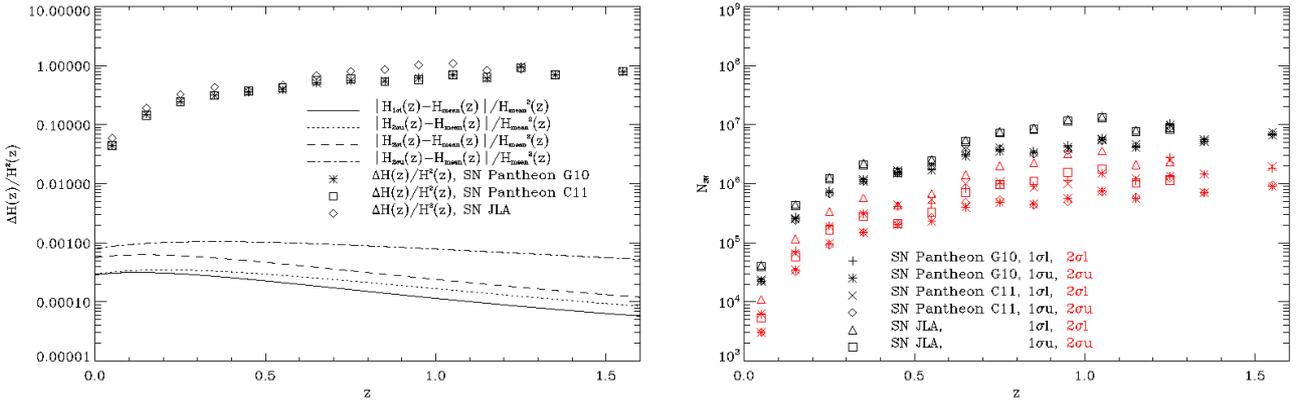

Fig. 4. Left: the theoretical relative differences $\Delta H_{model}(z)/H^2(z) \equiv |H_{1(2)\sigma}(z)-H_{mean}(z)|/H_{mean}^2(z)$ (for upper and lower limits) compared to $\Delta H(z)/H^2(z)$ from Supernovae compilations. Right: the minimal number of Supernovae that is necessary for distinguishing the models in left panel if the uncertainties of Supernovae magnitudes are the same as in the compilation from legend.

*Conclusion.*
We have found that despite the major increase in number of Supernovae in available compilations over the last 12 years the current prospects of using the dipole of luminosity distance for distinguishing the dark energy models are not bright. This is partly due to the fact that the uncertainties in determination of the cosmological parameters from other data are now much smaller and the tests for dark energy equation of state parameter should be more precise. Another reason is that now taken into account systematic errors result in the larger total ones. So, to make the luminosity distance dipole useful as the cosmological test it is necessary not only to increase largely the number of Supernovae (especially the low-redshift ones) in a dataset, but also to reduce the uncertainties of distance moduli by improving the light curve fitting and to control better the systematics.

*Acknowledgements*

This work has been supported in part by the Department of target training of Taras Shevchenko National University of Kyiv under National Academy of Sciences of Ukraine (project 6Ф). Author acknowledges the usage of CosmoMC package.

*References*
1. B. Javanmardi, C. Porciani, P. Kroupa, J. Pflamm-Altenburg, Probing the Isotropy of Cosmic Acceleration Traced By Type Ia Supernovae // Astrophys. J. **810**, 47 (2015).
2. U. Andrade, C.A.P. Bengaly, J.S. Alcaniz, B. Santos, Isotropy of low redshift type Ia supernovae: A Bayesian analysis // Phys. Rev. D **97**, 083518 (2018).
3. C. Bonvin, R. Durrer, M. Kunz, Dipole of the Luminosity Distance: A Direct Measure of H(z) // Phys. Rev. Lett. **96**, 191302 (2006).
4. C. Bonvin, R. Durrer, A. Gasparini, Fluctuations of the luminosity distance // Phys. Rev. D **73**, 023523 (2006).
5. B. Novosyadlyj, O. Sergijenko, R. Durrer, V. Pelykh, Quintessence versus phantom dark energy: the arbitrating power of current and future observations // J. Cosmol. Astropart. Phys. **06**, 042 (2013).
6. B. Novosyadlyj, O. Sergijenko, S. Apunevych, Distinguishability of scalar field models of dark energy with time variable equation of state parameter // J. Phys. Stud. **15**, 1901 (2011).
7. O. Sergijenko, B. Novosyadlyj, Scalar field as dark energy accelerating expansion of the Universe // Kin. Phys. Cel. Bod. **24**, p. 259-270 (2008).
8. A. Lewis, S. Bridle, Cosmological parameters from CMB and other data: a Monte Carlo approach // Phys. Rev. D **66**, 103511 (2002).
9. N. Jarosik, C.L. Bennett, J. Dunkley et.al., Seven-year Wilkinson Microwave Anisotropy Probe (WMAP) observations: sky maps, systematic errors and basic results // Astrophys. J. Suppl. **192**, 14 (2011).
10. E. Komatsu, K.M. Smith, J. Dunkley et.al., Seven-year Wilkinson Microwave Anisotropy Probe (WMAP) observations: cosmological interpretation // Astrophys. J. Suppl. **192**, 18 (2011).
11. D. Larson, J. Dunkley, G. Hinshaw et. al., Seven-year Wilkinson Microwave Anisotropy Probe (WMAP) observations: power spectra and WMAP-derived parameters // Astrophys. J. Suppl. **192**, 16 (2011).
12. W.J. Percival, B.A. Reid, D.J. Eisenstein et al., Baryon acoustic oscillations in the Sloan Digital Sky Survey Data Release 7 galaxy sample // Mon. Not. Roy. Astron. Soc. **401**, 2148 (2010).
13. A.G. Riess, L. Macri, S. Casertano et al., A Redetermination of the Hubble constant with the Hubble Space Telescope from a differential distance ladder // Astrophys. J. **699**, 539 (2009).
14. G. Steigman, Primordial nucleosynthesis in the precision cosmology era // Ann. Rev. Nucl. Part. Phys. **57**, 463 (2007).
15. E.L. Wright, Constraints on dark energy from supernovae, gamma ray bursts, acoustic oscillations, nucleosynthesis and large scale structure and the Hubble constant // Astrophys. J. **664**, 633 (2007).
16. R. Kessler, A.C. Becker, D. Cinabro et al., First-year Sloan Digital Sky Survey-II (SDSS-II) supernova results: Hubble diagram and cosmological parameters // Astrophys. J. Suppl. **185**, 32 (2009).
17. J. Guy, P. Astier, S. Baumont et al., SALT2: using distant supernovae to improve the use of type Ia supernovae as distance indicators // Astron. Astrophys. **466**, 11 (2007).
18. S. Jha, A.G. Riess, R.P. Kirshner, Improved distances to type Ia supernovae with multicolor light curve shapes: MLCS2K2 // Astrophys. J. **659**, 122 (2007).
19. Planck Collaboration, Planck 2015 results. XI. CMB power spectra, likelihoods, and robustness of parameters // Astron. Astrophys. **594**, A11 (2016).
20. Keck Array and BICEP2 Collaborations, BICEP2 / Keck Array VI: Improved Constraints On Cosmology and Foregrounds When Adding 95 GHz Data From Keck Array // Phys. Rev. Lett. **116**, 031302 (2016).
21. D. Parkinson et al., The WiggleZ dark energy survey: final data release and cosmological results // Phys. Rev. D **86**, 103518 (2012).
22. M. Betoule et al., Improved cosmological constraints from a joint analysis of the SDSS-II and SNLS supernova samples // Astron. Astrophys. **568**, A22 (2014).
23. G. Efstathiou, $H_0$ revisited // Mon. Not. Roy. Astron. Soc. **440**, 1138 (2014).
24. N. Suzuki, D. Rubin, C. Lidman et al., The Hubble space telescope cluster supernova survey: V. Improving the dark energy constraints above z > 1 and building an early-type-hosted supernova sample // Astrophys. J. **746**, 85 (2012).
25. D.M. Scolnic, D.O. Jones, A. Rest et al., The Complete Light-curve Sample of Spectroscopically Confirmed SNe Ia from Pan-STARRS1 and Cosmological Constraints from the Combined Pantheon Sample // Astrophys. J. **859**, 101 (2018).
26. M. Tanabashi et al. (Particle Data Group), Review of Particle Physics // Phys. Rev. D **98**, 030001 (2018).